\documentclass{iauc}
\usepackage{graphicx}

\title[Cosmology with Dwarf Spheroidals] %% give here short title %% 
{Near-Field Cosmology\\ with Local Group Dwarf Spheroidals}

\author[E.K.\ Grebel]   %% give here short author list %%
{Eva K.\ Grebel$^1$%
}

\affiliation{$^1$Astronomical Institute, University of Basel,
CH-4102 Binningen, Switzerland\break 
email: grebel@astro.unibas.ch}%\\[\affilskip]

\pubyear{2005}
\volume{198} %% insert here IAU Symposium No.
\pagerange{1--8}
\date{?? and in revised form ??}
\jname{Near-Field Cosmology With Dwarf Elliptical Galaxies}
\editors{H.\ Jerjen and B.\ Binggeli, eds.}
\begin{document}

\maketitle

\begin{abstract} The Local Group offers an excellent laboratory for
near-field cosmology by permitting us to use the resolved stellar content
of its constituent galaxies as probes of galaxy formation and evolution,
which in turn is an important means for testing cosmological models of
hierarchical structure formation.  In this review, we discuss the the least
massive, yet most
numerous type of galaxy in the Local Group, the dwarf spheroidal galaxies,
and compare their properties to cosmological predictions.  In particular,
we point out problems found with a simple building block scenario and with
effects expected from reionization.  We show that the star formation
histories of dSphs are inconsistent with the predicted cessation of star
formation after reionization; instead, extended star formation episodes are
observed.  The Galactic dSphs contain in part prominent intermediate-age
populations, whereas the Galactic halo does not.  Conversely, the M31
dSphs are almost entirely old, while the M31 halo contains a substantial
intermediate-age population.  These differences in the population 
structure as well as the differences in the modes of star formation
inferred from [$\alpha$/Fe] ratios make dSphs unlikely major contributors
to the build-up of the Galactic and M31 halo unless most of the accretion
occurred at early epochs.  On the other hand, there
is clear evidence for ongoing harassment and accretion of a number of
dSphs. 
%% add here a maximum of 10 keywords, to be taken form the file <Keywords.txt>.

\end{abstract}

\firstsection % if your document starts with a section,
              % remove some space above using this command.
\section{Introduction}

The Local Group is an excellent laboratory for studies of galaxy evolution
at the highest possible resolution in that it provides us with a wide range
of different galaxy types and a variety of environments.  Yet the Local
Group is a poor group, contains relatively few galaxies, and lacks very
massive galaxies, such as large ellipticals.  Similar to many other nearby
groups, the mass and the luminosity of the Local Group are dominated by two
large spirals, the Milky Way and M31.  Most of the other Local Group
members are dwarf galaxies, and the majority of them are found in close
proximity to the two large spirals.

Dwarf galaxies are often considered building blocks of more massive
galaxies in models of hierarchical structure formation.  Dwarf galaxies
come in many different flavors and cover a range of masses, luminosities,
morphologies, gas content, star formation histories, etc.  The distinction
between dwarf galaxies and larger galaxies is somewhat fuzzy.  The
difference is primarily a luminosity difference -- it is customary to call
galaxies with absolute magnitudes of $M_V > -18$ dwarf galaxies.  Gas-rich
dwarfs include dwarf spirals, dwarf irregulars (dIrrs) and blue compact
dwarf galaxies, which usually show differing levels of ongoing star
formation.  Gas-poor dwarfs are primarily dwarf ellipticals (dEs).  These
can be further subdivided into subtypes such as the  more massive, strongly
centrally concentrated dwarf ellipticals with higher surface brightness,
and the less massive, faint, fairly diffuse dwarf spheroidals (dSphs) (see
also Gallagher \& Wyse 1994; Grebel, Gallagher, \& Harbeck 2003).  

What makes the Local Group special (apart from its being our home) is that
here we have the possibility to actually resolve its constituent galaxies
into individual stars and to study the properties of these stars.  We can
use these stars as probes of the past -- they permit us to uncover the
evolutionary histories of their host galaxies.  Moreover, they permit us to
study these evolutionary histories at a level of detail and accuracy that
is unmatched by any more distant galaxy, where only the integrated light
can be studied.  The Local Group is the only place where we can analyze
even {\em ancient}\ stars and uncover the {\em early}\ formation history of
individual galaxies beyond our own, particularly of the dwarf companions of
the Milky Way.  To summarize, the Local Group is ideally suited for studies
of ``near-field cosmology'', i.e., for studies of galaxy evolution over
cosmological epochs based on their resolved stellar fossil record, and for
tests of the corresponding cosmological models.  

\section{Local Group Dwarf Spheroidals}

The galaxy census of the Local Group remains uncertain.  Within the Local
Group's volume as defined by its zero velocity surface of $\sim$ 1 Mpc
(Karachentsev et al.\ 2002), we currently know of some 38 probable member
galaxies.  Some were only recently discovered, and additional faint
candidates continue to be found (e.g., Zucker et al.\ 2004a,b).  All of the
newly discovered dwarfs are dSphs, the least massive, least luminous
galaxies known,  and thus contribute to the faint end of the galaxy
luminosity function.  For reviews on Local Group dwarfs, see, e.g., Grebel
(1997, 1999, 2000, 2001, 2005), Mateo (1998), and van den Bergh (1999,
2000). 

DSphs are usually the most numerous type of galaxy in galaxy groups and are
characterized by M$_V < -14$ mag, $\mu_V < 22$ mag arcsec$^{-2}$, $M_{\rm
HI} < 10^5$ M$_{\odot}$, and $M_{tot} \sim 10^7$ M$_{\odot}$.  Often their
stellar populations are purely old, but mixtures of old and
intermediate-age populations are found as well.  In dSphs where several
populations can be distinguished the younger and/or more metal-rich
populations are more centrally concentrated, indicating extended star
formation episodes in the centers of the shallow potential wells of their
parent galaxies (Harbeck et al.\ 2001).  The gas deficiency of dSphs
remains an unsolved puzzle -- dSphs typically contain even less gas than
expected from red giant mass loss over time scales of several Gyr.  The
metallicity--luminosity relations of dSphs and dIrrs show the usual trend
of increasing metallicity with increasing galaxy luminosity, but the
relations are offset from each other: DSphs have higher mean stellar
metallicities at a given optical luminosity, which may indicate more rapid
star formation and enrichment at early times as compared to dIrrs (Grebel
et al.\ 2003).

DSphs do not seem to be  supported by rotation and appear to contain large
amounts of dark matter.  The latter is inferred from the high velocity
dispersion and the resulting high mass-to-light ratios derived under the
assumption of virial equilibrium.  Indirectly, a high dark matter content
is also supported by the morphology of some nearby dSphs (Odenkirchen et
al.\ 2001) and by the observed lack of a significant depth extent (Klessen,
Grebel, \& Harbeck 2003).  The radial velocity dispersion profiles of dSphs
fall off large radii (Wilkinson et al.\ 2004), possibly indicating the
presence of a kinematically cold stellar population at the outermost radii. 

If dwarf galaxies in general and dSphs in particular are indeed building
blocks of larger galaxies, then today's dwarf population may be considered
to be the surviving population of satellites that has not yet been
accreted.  The most numerous type of dwarfs in galaxy groups, the dSphs,
may then be the most pristine members of the original building block
population.  Studying dSphs may teach us about the properties of objects
that presumably were once accreted in large numbers to form galaxies like
the Milky Way.  Alternatively, it is conceivable that dSphs are in fact not
fossil building blocks, but stripped remnants of disrupted and originally
much more massive galaxies that have since merged.  To find out more about
the nature of dSphs and their cosmological significance we need to
understand their past and present-day properties.

\section{Dwarf Spheroidals:  The Earliest Measurable Epoch of Star Formation
and Its Cosmological Implications}

Cold dark matter models predict that low-mass systems were the first sites
of star formation, possibly as early as at a redshift of 30 (e.g., Barkana
\& Loeb 2001).  Since larger systems form through hierarchical merging of
smaller systems, they should contain surviving populations of these early
epochs of star formation.  Furthermore, several models predict that small
galaxies should have formed most of their stars prior to reionization,
while reionization would have suppressed further star formation activity.
In fact, galaxies less massive than $10^9$ M$_{\odot}$ should have lost
their star-forming material during reionization (e.g., Susa \& Umemura
2004).  Hence one would expect that low-mass galaxies contain ancient
populations, while star formation should have ceased after reionization.

These are predictions that can be tested in the dwarfs in our immediate
surroundings.  The least massive dwarfs, the dSphs, should have been most
severely affected.  Deep color-magnitude diagram data of these dwarfs that
reach below the oldest main-sequence turn-offs permit us to carry out
relative age dating of their old populations with internal accuracies of
fractions of $\sim$1~Gyr, the highest accuracy currently attainable for any
method for old stars.  Note that this method can only be applied to
populations sufficiently numerous to form detectable main-sequence
turn-offs.  This holds only for old Population II stars, while potential
Population III stars are far too few even in our Milky Way.  The
differential ages of old populations in dwarf galaxies -- either field
populations or globular clusters -- can then be compared to the ages of the
oldest Galactic globular clusters of similar composition.

\begin{figure}
 \includegraphics[height=3in,width=5.3in]{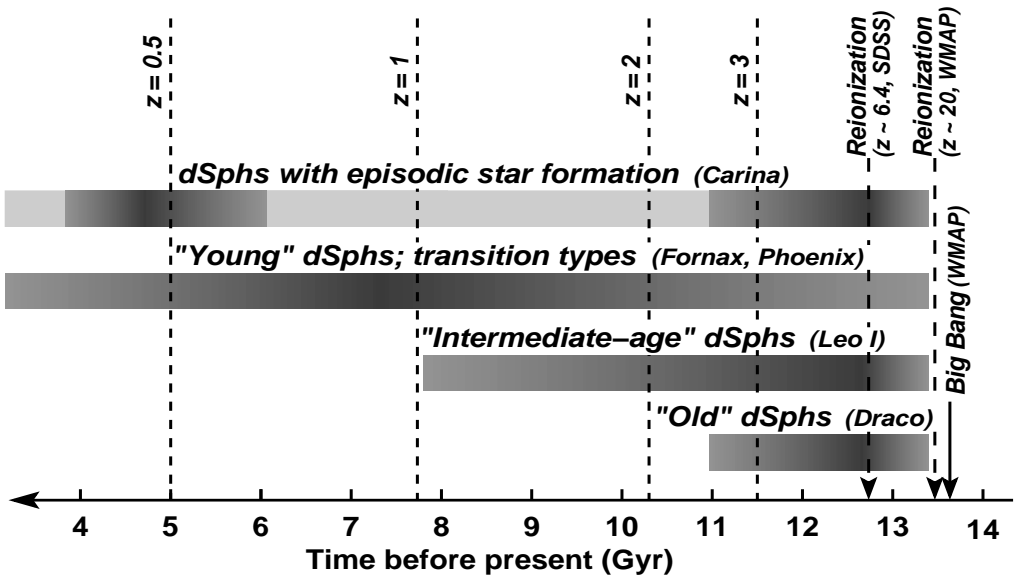}
  \caption{
Sketch indicating the approximate duration of star formation
episodes in dSph galaxies ($\sim 10^7$ M$_{\odot}$).  The 
adopted beginning and end of the reionization epoch are based on
results from WMAP and from the Sloan Digital Sky Survey.  The 
predicted cessation of star formation after reionization 
is not observed.  For details, see Grebel \& Gallagher (2004).
    }
\end{figure}

Importantly, this method reveals that (1) old populations are ubiquitous
(but their fractions vary) and (2) the oldest ages in all of the galaxies
studied so far are indistinguishable within the measurement accuracy (see
Grebel 2000 and Grebel \& Gallagher 2004 for details).  All nearby dwarf
galaxies studied in sufficient detail were shown to contain ancient
populations.  Moreover, these nearby Milky Way companions and the Milky Way
itself share a common epoch of star formation for their ancient Population
II (within $\sim 1$ Gyr).  These observations are consistent with the
expectations from the building block scenario. 

However, the predicted cessation of star formation after reionization,
expected to have affected particularly dSphs owing to their low mass, is
not observed (Grebel \& Gallagher 2004).  Instead, even dSphs entirely
dominated by old populations show evidence for star formation extending
over many Gyr (Harbeck et al.\ 2001; Ikuta \& Arimoto 2002).  In dSphs with
a mixture of populations we usually find star formation episodes that
lasted many Gyr without being interrupted by a pronounced, long hiatus
to the extent that we can measure the duration of star formation
(exception: Carina with its episodic star formation).  This may mean
that the above quoted cosmological models are incorrect and do not properly
consider other effects that might prevent complete photoevaporation (e.g.,
Susa \& Umemura 2004).  On the other hand, photoionization is one plausible
way to circumvent the substructure problem (e.g., Somerville 2002).
Alternatively, it is conceivable that dSphs were once considerably more
massive (by roughly a factor of 100), which could also have prevented
photoionization squelching.  In this case the galaxies observed today as
dSphs must have undergone substantial mass loss. 

\section{Dwarf Spheroidal Star Formation Histories and Abundances}

Thanks to deep, high-resolution photometry and synthetic color-magnitude
diagram techniques, fairly detailed knowledge of the star formation
histories of Local Group dSphs is now available.  The resulting picture is
one of high complexity:  No two dSphs exhibit the same star formation
history (Grebel 1997).  As mentioned already, all dSphs studied in detail
so far were found to contain old Population II stars.  Some dSphs are
dominated by ancient stars, others only have a small old population and a
dominant intermediate-age population, and there is one example of a dSph
that experienced star formation as recently as a few hundred Myr ago
(Fornax, see Grebel \& Stetson 1999).  Generally, star formation has
proceeded continuously in these galaxies, although the amplitude varied and
eventually declined at intermediate or younger ages (e.g., Grebel et al.\
2003).  Only one dSph with clearly episodic star formation is known
(Carina, Smecker-Hane et al.\ 1994 and Monelli et al.\  2003).  DSphs with
several populations typically show population gradients in the sense that
more metal-rich and/or younger populations are more centrally concentrated
(Harbeck et al.\ 2001).  Substructure of this kind is not necessarily
symmetrically distributed (e.g., Stetson et al.\ 1998).  

While the past decade was mainly one of photometrically derived star
formation histories, we are now entering an era where the age-metallicity
degeneracy, which is inherent to purely photometric determinations, can be
broken by adding spectroscopic abundance information (e.g., Tolstoy et al.\
2001; Pont et al.\ 2004, Cole et al.\ 2005, Koch et al. in these
proceedings).   This will ultimately permit us to derive detailed
age-metallicity relations for these galaxies.

\subsection{Comparing stellar populations and star formation histories}

Comparing star formation histories of dwarf galaxies in general and dSphs
in particular (e.g., Grebel 1997, 1999), one finds variations in the
duration of star formation, in the star formation rates as a function of
time, and in the enrichment.  In spite of being overall metal-poor, all
dSph galaxies that were studied spectroscopically so far show a spread of
metallicities of typically 1 dex in [Fe/H] or more (e.g., Shetrone et al.\
2001, 2003; Bonifacio et al.\ 2004).  There appears to be a trend of
increasing intermediate-age population fractions with increasing distance
from the Milky Way among the Galactic dSphs (van den Bergh 1994; Grebel
1997), which may be due to the environmental impact of the Milky Way.  

If environment was indeed the governing factor determining the evolution of
these low-mass galaxies, then one should expect to find a similar trend
among the dSph companions of M31.  However, this is not observed.  Although
M31's dSphs cover a comparable range of distances as their Galactic
counterparts, they all appear to be dominated by old populations and lack
the indicators of prominent intermediate-age populations present in the
more distant Milky Way dSphs (Harbeck et al.\ 2001, 2004, 2005). 

Considering what we can infer from present-day dSphs about their star
formation histories, how do they fit in as potential building blocks?  With
respect to stellar populations, dSphs dominated by old populations are
compatible with the stellar content of the Galactic halo.  DSphs with
substantial intermediate-age populations seem less likely to have made a
major contribution to the build-up of the halo of our Milky Way (Unavane et
al.\ 1996).  On the other hand, this problem would be diminished if most of
the minor merger events took place at very early epochs.   Comparing the
old, metal-poor stellar populations in M31's dSphs to M31's halo indicates
that the dSphs cannot have been primary building blocks of M31's halo since
it was found to contain a substantial contribution from intermediate-age,
comparatively metal-rich populations (Brown et al.\ 2003).   An old,
metal-poor halo population, however, has been detected as well (Brown et
al.\ 2004), and again the population differences would be less severe if
most of the dSph accretion had taken place at very early epochs, whereas
the remainder of the younger halo of M31 would have been formed through the
later accretion of more massive and more evolved galaxies.  -- These
statements assume that dSphs have not changed appreciably over time (e.g.,
did not lose substantial amounts of mass) and that their observed stellar
content permits one to arrive at a fair representation of their
evolutionary history. 

\begin{figure}
 \includegraphics[height=3.4in,width=5.3in]{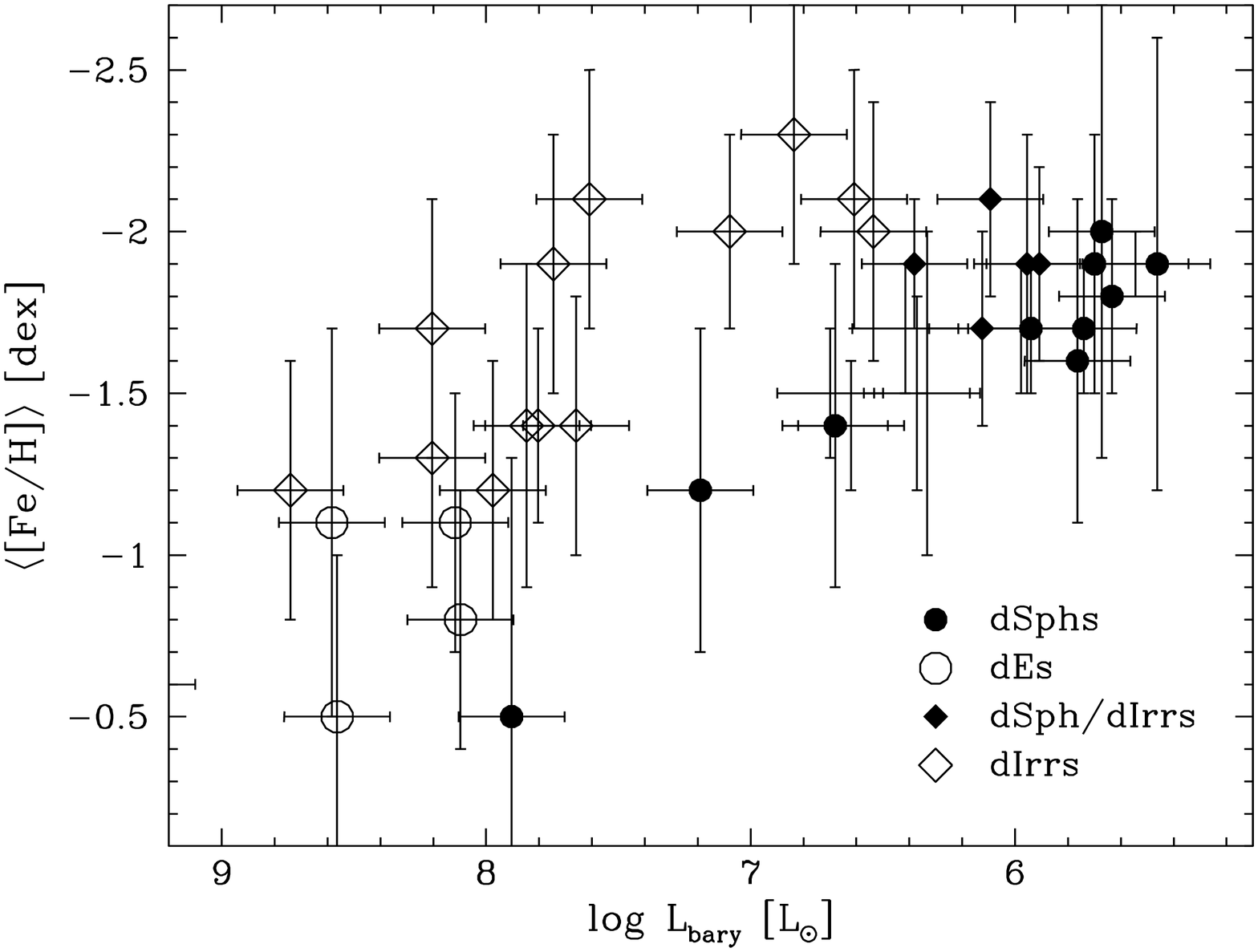}
  \caption{
Mean stellar metallicity of Population II stars versus baryonic luminosity
for different classes of dwarf galaxies as indicated in the legend.  Note
the offset between gas-deficient (filled symbols) and gas-rich (open
symbols) dwarfs.  At the same galaxy luminosity, the old populations of 
dSphs are more metal-rich than those of dIrrs.  Thus in contrast to dIrrs,
dSphs must have experienced comparatively rapid early enrichment.  Note the
location of the so-called dIrr/dSph transition-type galaxies, which
combine population properties of dSphs with ongoing star formation and a
measurable gas content in the diagram.  Their properties make them
plausible dSph progenitors.
For details, see Grebel, Gallagher, \& Harbeck (2003).
    }
\end{figure}

\subsection{Are DSph Abundance Patterns Consistent with the Building Block
Scenario?}

During the past few years more and more detailed, high-resolution abundance
analyses of individual red giants in dSphs have become available, leading
to a growing, yet still limited body of knowledge about their detailed
element abundance ratios.  In particular, [$\alpha$/Fe] ratios, r- and
s-process abundances are being measured.  

If dSphs were dominant contributors to the build-up of the Galactic halo,
their abundance patterns should match those of the halo.  However, the
existing measurements show pronounced differences to the abundance ratios
in our Galactic halo: Dwarfs are characterized by slower star formation
rates, leading to reduced [$\alpha$/Fe] ratios at lower metallicity
([Fe/H]) than found in the Galactic halo.  This can be interpreted as a
signature of a larger contribution of supernovae of Type Ia early on, such
that a solar [$\alpha$/Fe] is reached sooner (e.g., Matteucci 2003).   In
contrast, the Galactic halo experienced comparatively rapid star formation
accompanied by gas removal, leading to low metallicities with higher
$\alpha$ element ratios.  These different properties lead to the conclusion
that dSphs cannot have contributed in a major way to the build-up of the
Galactic halo (Shetrone et al.\ 2001), unless the majority of the minor
merger events occurred at very early epochs when the abundance ratios in
the Milky Way and in the dSphs were still very similar. 

\subsection{Morphological segregation and the metallicity-luminosity relation for dwarf galaxies}

Going back to global metallicities (``[Fe/H]''), what can these tell us
about galaxy evolution and interrelations between different galaxies?  Now
we do not consider dSphs as building blocks of larger galaxies, but dSphs
as the stripped remnants of initially more massive galaxies.  Clearly,
dSphs must once have been more massive and considerably more gas-rich in
order to have formed the stars we observe in them today.  Their present-day
gas deficiency still lacks a satisfactory explanation (Gallagher et al.\
2003, Grebel et al.\ 2003).  What were the progenitors of dSphs?  The first
type of galaxies that comes to mind are dIrrs, gas-rich, irregularly shaped
dwarfs with ongoing star formation yet also with very old populations.  

Could dSphs simply be stripped dIrrs?  Taken at face value, the
morphological segregation observed in the Local Group (as well as in other
groups) would seem to support this idea (see, e.g., Fig.\ 1 in Grebel
1995): Gas-deficient dwarf galaxies (dEs, dSphs) are usually found within
300 kpc around more massive galaxies, while gas-rich dwarfs (esp.\ dIrrs)
are also (and predominantly) found at larger distances.  When plotting
distance from the nearest primary vs. H\,{\sc i} content, there is a clear
tendency to find dSphs with H\,{\sc i} mass limits below $10^5$ M$_{\odot}$
within 300 kpc, while dIrr galaxies with typical H\,{\sc i} masses $> 10^7$
M$_{\odot}$ tend to lie at distances $> 400$ kpc (Grebel et al.\ 2003,
their Fig.\ 3).  The proximity to massive galaxies and interactions with
these may be an efficient agent in removing material from the dwarfs (e.g.,
Mayer et al.\ 2001).  

On the other hand, the luminosity-metallicity relations of dSphs and dIrrs
have long been known to differ.  While for both classes of galaxies the
metallicity increases with luminosity (and hence with mass), the two
relations are offset from one another (e.g., Skillman \& Bender 1995) in
the sense that dSphs are more metal-rich for a given luminosity.  However,
the luminosity-metallicity relations are based on different tracers:  For
dIrrs, usually the present-day oxygen abundances as measured in H\,{\sc ii}
regions are used, while for dEs and dSphs, metallicities of old populations
(and occasionally oxygen abundances of intermediate-age planetary nebulae)
are used.  Thus the metallicities of populations of very different ages as
well as nebular abundances versus stellar abundances are compared.  

This mixture of different evolutionary stages and different metallicity
indicators is unsatisfactory.  Therefore we decided to attempt to compare
apples with apples:   In order to compare not only mean {\em stellar
metallicities}\ in dIrrs and dSphs, but also the metallicities of the {\em
same populations} (i.e., of stars of similar age), we chose  old Population
II giants, which are found  in all LG dwarf galaxies.  We used (1) {\em old
red giants} in dSphs and in the outskirts of dIrrs (where old populations
dominate), (2) {\em spectroscopic abundances} wherever available (from our
own and literature data), and (3) {\em photometric abundances} elsewhere.
The resulting data set may not yet have an ideal degree of homogeneity, but
is the best and most comprehensive one currently available (Grebel et al.\
2003).  In the coming years, undoubtedly stellar spectroscopic measurements
will also become available for those dwarfs for which we only have
photometric estimates at present.

Interestingly, even when confining the comparison of luminosity-metallicity
relations to old populations, the differences continue to exist.  {\em Thus
at the same galaxy luminosity, the old populations of dSphs are more
metal-rich than those of dIrrs}.  This indicates that in contrast to dIrrs,
dSphs must have experienced fairly rapid early enrichment (Grebel et al.\
2003).  This and several other factors make dIrrs unlikely progenitors of
dSphs.  If dSphs are stripped remnants of more massive galaxies, then the
fact that they do follow a baryonic luminosity-metallicity relation
indicates that they must have continued to form stars and to experience
enrichment even after the main mass removal occurred.  Grebel et al.\
(2003) present a series of arguments why dIrr/dSph transition-type galaxies
appear to be fairly plausible dSph progenitors, suggesting a gentle and
slow transition from one kind of low-mass galaxy to another.

\section{Harassment and accretion}

While we presented arguments against a simple cosmological building block
scenario in the preceding paragraphs, there is clear evidence for ongoing
harassment and accretion of dwarf galaxies.  The most prominent examples
of ongoing accretion are the tidal streams of the Sagittarius dSph galaxy
(Ibata et al.\ 1994), and the giant stream of metal-rich giants around M31
(Ibata et al.\ 2001).  Additional stellar overdensities have been detected
in the Milky Way, e.g., the Monoceros feature (Newberg et al.\ 2002; Yanny
et al.\ 2003), the Canis Major overdensity (Martin et al.\ 2004), the
Triangulum-Andromeda feature (Rocha-Pinto et al.\ 2004) and more
substructure near M31 (Zucker et al.\ 2004a).  These may be parts of the
tidal tails of disrupted dwarfs.  The continuation of deep wide-field
surveys and the addition of spectroscopic data for phase-space information
will ultimately permit us to identify and constrain less pronounced
accretion events and their number, providing an important observable for
hierarchical structure formation. 

Evidence for harassment is apparent in the S-shaped surface density
profile of the Galactic dSph Ursa Minor (Palma et al.\ 2003) and in the
twisted isophotes of the M31 dE companions M32 and NGC\,205 (Choi et al.\
2002).   These and other dSphs may eventually be accreted as well.  A
crucial bit of information in this context is the knowledge of the orbits
of dwarf companions -- something the Gaia mission will help to establish. 

\section{Concluding remarks}

What is the cosmological role of dSphs?  With regard to the oldest
measurable ages and the earliest epoch of star formation, we find
consistency with expectations from cosmological modes.  There appears to be
a common epoch of early star formation in the Milky Way and in its dwarf
companions.  In contrast to model predictions, the expected cessation of
star formation after reionization is, however, not observed in dSphs.
The observed population structure in the (very different)
halos of M31 and of the Milky Way, makes it seem unlikely that
(present-day) dSphs played a major role in the build-up of the halo of the
two spirals.  The large variations in the star formation histories of dSphs
and the presence of younger populations than in the Galactic halo can be
reconciled with the building block scenario if most of the accretion events
occurred very early on.  

The global metallicities of the Milky Way dSphs are well-matched to those
observed in the Galactic halo, but this is not the case for the M31 dSph
companions.  With regard to detailed chemical element abundance ratios, it
is emerging that dSphs cannot have been dominant contributors to halo
build-up unless -- again -- the merger events would have taken place at
very early times.  The differences in the metallicity-luminosity relation
of different types of dwarfs seem to exclude that dSphs are simply
stripped dIrrs.

Both disruption and accretion of dwarf companions are still occurring
today, demonstrating that dSphs must have played some role in the growth of
larger galaxies.  Unfortunately, the number and importance of accretion
events remains unclear for either of the two large spirals in the Local
Group.  

In spite of admirable progress, dwarfs remain an evolutionary puzzle.  They
are excellent and important test cases of cosmological predictions.
Regardless of their cosmological importance, however, dwarf galaxies are
also interesting in their own right!

\begin{acknowledgments} Many thanks to Helmut Jerjen and Bruno Binggeli for
a wonderful conference and for their patience while this contribution was
finished.  I am also indebted to Jay Gallagher for a critical reading of
this text.  
\end{acknowledgments}


\begin{thebibliography}{}

\bibitem[Barkana \& Loeb (2001)]{Bark01} {Barkana, R., \& Loeb, A.} 1999,
\textit{Phys.\ Rep.}, {349}, 125

\bibitem[Bonifacio \etal\ (2004)]{Boni04} {Bonifacio, P., 
Sbordone, L., Marconi, G., Pasquini, L., \& Hill, V.} 2004, \textit{A\&A}, 
414, 503

\bibitem[Brown \etal\ (2003)]{Brown03} {Brown, T.~M., Ferguson, 
H.~C., Smith, E., Kimble, R.~A., Sweigart, A.~V., Renzini, A., Rich, R.~M., 
\& VandenBerg, D.~A.} 2003, \textit{ApJ}, 592, L17 

\bibitem[Brown \etal\ (2004)]{Brown04} {Brown, T.~M., Ferguson, 
H.~C., Smith, E., Kimble, R.~A., Sweigart, A.~V., Renzini, A., \& Rich, 
R.~M.} 2004, \textit{AJ}, {127}, 2738 

\bibitem[Choi \etal\ (2002)]{Choi02} {Choi, P.~I., Guhathakurta, 
P., \& Johnston, K.~V.} 2002, \textit{AJ}, 124, 310 

\bibitem[Cole \etal\ (2005)]{Cole05} {Cole, A.~A., Tolstoy, E., 
Gallagher, J.~S., \& Smecker-Hane, T.~A.} 2005, \textit{AJ}, 129, 1465 

\bibitem[Gallagher \& Wyse (1994)]{Gall94} {Gallagher, J.~S., \& 
Wyse, R.~F.~G.} 1994, \textit{PASP}, 106, 1225

\bibitem[Gallagher \etal\ (2003)]{Gall03} {Gallagher, J.~S., 
Madsen, G.~J., Reynolds, R.~J., Grebel, E.~K., \& Smecker-Hane, T.~A.} 
2003, \textit{ApJ}, 588, 326

\bibitem[Grebel (1997)]{Grebel97}
{Grebel, E.~K.} 1997, \textit{Reviews in Modern Astronomy}, {10}, 27

\bibitem[Grebel (1999)]{Grebel99}
{Grebel, E.~K.} 1999, in: P.\ Whitelock \& R.\ Cannon (eds.), 
\textit{The Stellar Content of the Local Group}, IAU Symp.\ 192, 
(San Francisco: ASP), p.\ 17

\bibitem[Grebel (2000)]{Grebel00}
{Grebel, E.~K.} 2000, in: F.\ Favata, A.A.\ Kaas, \& A.\ Wilson (eds.),
\textit{Star formation from the small to the large scale},
33rd ESLAB Symposium, ESA-SP 445 (Noordwijk: ESA), p.\ 87

\bibitem[Grebel (2001)]{Grebel01}
{Grebel, E.~K.} 2001, \textit{A\&SSS}, {277}, 231

\bibitem[Grebel (2005)]{Grebel05}
{Grebel, E.~K.} 2005, in: J.\ Mikolaewska \& A.\ Olech (eds.),
\textit{Stellar Astrophysics with the World's Largest Telescopes},
AIP Conf.\ Proc.\ Vol.\ 752 (New York: AIP), p.\
161 

\bibitem[Grebel, Gallagher \& Harbeck (2003)]{GreGalHar03}
{Grebel, E.~K., Gallagher, J.~S., \& Harbeck, D.} 2003, \textit{AJ}, {125},
1926

\bibitem[Grebel \& Gallagher (2004)]{GreGal04}
{Grebel, E.~K., \& Gallagher, J.~S.} 2004, \textit{ApJ}, {610}, L89

\bibitem[Grebel \& Stetson (1999)]{GreSte99}
{Grebel, E.~K., \& Stetson, P.~B.} 1999, in: P.\ Whitelock \& R.\ Cannon (eds.),
\textit{The Stellar Content of the Local Group}, IAU Symp.\ 192,
(San Francisco: ASP), p.\ 165

\bibitem[Harbeck \etal\ (2001)]{Harb01} {Harbeck, D.} et al.\ 
2001, \textit{AJ}, {122}, 3092 

\bibitem[Harbeck \etal\ (2004)]{Harb04}
{Harbeck, D., Gallagher, J.~S., \& Grebel, E.~K.} 2004, \textit{AJ}, {127},
2711

\bibitem[Harbeck \etal\ (2005)]{Harb05}
{Harbeck, D., Gallagher, J.~S., Grebel, E.~K., Koch, A., \& Zucker, D.~B.}
2005, \textit{ApJ}, 623, 159 

\bibitem[Ikuta \& Arimoto (2002)]{Ikuta02} {Ikuta, C., \& Arimoto, N.}
2002, \textit{A\&A}, {391}, 55

\bibitem[Ibata \etal\ (1994)]{Ibata94} {Ibata, R.~A., Gilmore, 
G., \& Irwin, M.~J.} 1994, \textit{Nature}, 370, 194 

\bibitem[Ibata \etal\ (2001)]{Ibata01} {Ibata, R., Irwin, M., 
Lewis, G., Ferguson, A.~M.~N., \& Tanvir, N.} 2001, \textit{Nature}, 412, 49 

\bibitem[Karachentsev \etal\ (2002)]{Kara02} {Karachentsev, I.~D., 
et al.} 2002, \textit{A\&A}, {389}, 812 

\bibitem[Klessen \etal\ (2003)]{Klessen03} {Klessen, R.~S., Grebel, 
E.~K., \& Harbeck, D.} 2003, \textit{ApJ}, {589}, 798 

\bibitem[Martin et al.(2004)]{Martin04} {Martin, N.~F., Ibata, 
R.~A., Bellazzini, M., Irwin, M.~J., Lewis, G.~F., \& Dehnen, W.} 2004, 
\textit{MNRAS}, 348, 12 

\bibitem[Mateo (1998)]{Mateo98}
{Mateo, M.} 1998, \textit{ARA\&A}, {36}, 435

\bibitem[Matteucci (2003)]{Matte03} {Matteucci, F.} 2003, \textit{Ap\&SS},
{284}, 539

\bibitem[Mayer \etal\ (2001)]{Mayer01} {Mayer, L., Governato, F., 
Colpi, M., Moore, B., Quinn, T., Wadsley, J., Stadel, J., \& Lake, G.} 
2001, \textit{ApJ}, 547, L123 

\bibitem[Monelli \etal\ (2003)]{Mone03} {Monelli, M., et al.} 
2003, \textit{AJ}, 126, 218 

\bibitem[Newberg \etal\ (2002)]{Newberg02} {Newberg, H.~J., et al.} 
2002, \textit{ApJ}, 569, 245 

\bibitem[Odenkirchen \etal\ (2001)]{Odenk01} {Odenkirchen, M., et 
al.} 2001, \textit{AJ}, {122}, 2538

\bibitem[Palma \etal\ (2003)]{Palma03} {Palma, C., Majewski, 
S.~R., Siegel, M.~H., Patterson, R.~J., Ostheimer, J.~C., \& Link, R.} 
2003, \textit{AJ}, 125, 1352

\bibitem[Pont \etal\ (2004)]{Pont04} {Pont, F., Zinn, R., 
Gallart, C., Hardy, E., \& Winnick, R.} 2004,  \textit{AJ}, 127, 840

\bibitem[Rocha-Pinto \etal\ (2004)]{Rocha04} {Rocha-Pinto, H.~J., 
Majewski, S.~R., Skrutskie, M.~F., Crane, J.~D., \& Patterson, R.~J.} 2004, 
\textit{ApJ}, 615, 732 


\bibitem[Shetrone \etal\ (2001)]{Shet01}
{Shetrone, M.~D., C\^ot\'e, P., \& Sargent, W.~L.~W.} \textit{ApJ}, 2001,
{548}, 592

\bibitem[Shetrone \etal\ (2003)]{Shet03}  {Shetrone, M., Venn, 
K.~A., Tolstoy, E., Primas, F., Hill, V., \& Kaufer, A.} 2003, 
\textit{AJ}, {125}, 684

\bibitem[Skillman \& Bender (1995)]{Skill95} Skillman, E.~D., \& 
Bender, R.\ 1995, \textit{RMxAC}, 3, 25 

\bibitem[Smecker-Hane \etal\ (1994)]{Sme94} {Smecker-Hane, 
T.~A., Stetson, P.~B., Hesser, J.~E., \& Lehnert, M.~D.} 1994, 
\textit{AJ}, {108}, 507 

\bibitem[Somerville(2002)]{Somer02} {Somerville, R.~S.} 2002, 
\textit{ApJ}, {572}, L23 

\bibitem[Stetson \etal\ (1998)]{Stetson98} {Stetson, P.~B., Hesser, 
J.~E., \& Smecker-Hane, T.~A.} 1998, \textit{PASP}, {110}, 533

\bibitem[Susa \& Umemura (2004)]{Susa04}
{Susa, H., \& Umemura, M.} 2004, \textit{ApJ}, {600}, 1

\bibitem[Tolstoy \etal\ (2001)]{Tolstoy01} {Tolstoy, E., Irwin, 
M.~J., Cole, A.~A., Pasquini, L., Gilmozzi, R., \& Gallagher, J.~S.} 2001, 
\textit{MNRAS}, {327}, 918 

\bibitem[Unavane \etal\ (1996)]{Una96} {Unavane, M., Wyse, R.~F.~G., 
\& Gilmore, G.} 1996, \textit{MNRAS}, 278, 727 

\bibitem[van den Bergh (1999)]{vdB99}
{van den Bergh, S.} 1999, \textit{A\&ARv}, {9}, 273

\bibitem[van den Bergh (2000)]{vdB00}
{van den Bergh, S.} 2000, \textit{The Galaxies of the Local Group},
Cambridge Astrophysics Series Vol.\ 35 (Cambridge: Cambridge University Press)

\bibitem[Wilkinson \etal\ (2004)]{Wilk04} {Wilkinson, M.~I., 
Kleyna, J.~T., Evans, N.~W., Gilmore, G.~F., Irwin, M.~J., \& Grebel, 
E.~K.} 2004, \textit{ApJ}, {611}, L21

\bibitem[Yanny \etal\ (2003)]{Yanny03} {Yanny, B., et al.} 2003, 
\textit{ApJ}, 588, 824 

\bibitem[Zucker \etal\ (2004a)]{Zucker04a} {Zucker, D.~B., et al.\ }            
2004a, \textit{ApJ}, {612}, L117

\bibitem[Zucker \etal\ (2004b)]{Zucker04b} {Zucker, D.~B., et al.\ }
2004b, \textit{ApJ}, {612}, L121

\end{thebibliography}
\end{document}